\documentclass[12pt]{article}
\usepackage{amsmath}
\def\a{\alpha}
\def\b{\beta}

\def\g{\gamma}
\def\p{\psi}

\def\t{\theta}

\def\s{\sigma}

\def\D{\Delta}
\def\G{\Gamma}

\def\be{\begin{equation}}
\def\ee{\end{equation}}
\def\arr{\begin{array}{rll}}
\def\ea{\end{array}}
\def\bea{\begin{eqnarray}}
\def\eea{\end{eqnarray}}

\def\N2{$N{=}2$}

\def\>{\rangle}
\def\<{\langle}
\def\+{\dagger}
\def\={\ =\ }

\begin{document}
\renewcommand{\thefootnote}{\fnsymbol{footnote}}
\begin{titlepage}
\setcounter{page}{0}
\begin{flushright}
LMP-TPU--7/10  \\
\end{flushright}
\vskip 1cm
\begin{center}
{\LARGE\bf  Particle dynamics near extreme}\\
\vskip 0.5cm
{\LARGE\bf     Kerr throat and supersymmetry }\\
\vskip 2cm
$
\textrm{\Large Anton Galajinsky\ }
$
\vskip 0.7cm
{\it
Laboratory of Mathematical Physics, Tomsk Polytechnic University, \\
634050 Tomsk, Lenin Ave. 30, Russian Federation} \\
{Email: galajin@mph.phtd.tpu.ru}

\end{center}
\vskip 1cm
\begin{abstract} \noindent
The extreme Kerr throat solution is believed to be non--supersymmetric.
However, its isometry group $SO(2,1) \times U(1)$ matches precisely the bosonic subgroup
of $\mathcal{N}=2$ superconformal group in one dimension. In this paper we construct $\mathcal{N}=2$ supersymmetric
extension of a massive particle moving near the horizon of the extreme Kerr black hole.
Bosonic conserved charges are related to Killing vectors
in a conventional way. Geometric interpretation of
supersymmetry charges remains a challenge.

\end{abstract}

\vspace{0.5cm}

PACS: 04.70.Bw; 11.30.-j; 11.30.Pb \\ \indent
Keywords: extreme Kerr throat geometry, conformal mechanics, $\mathcal{N}=2$ supersymmetry
\end{titlepage}

\renewcommand{\thefootnote}{\arabic{footnote}}
\setcounter{footnote}0

\noindent
{\bf 1. Introduction}\\

In 1999, Bardeen and Horowitz \cite{bh} derived the near horizon limit of the extreme Kerr solution and found that
the isometry group is enhanced to $SO(2,1) \times U(1)$. Because the first factor is the conformal group in one dimension,
it was conjectured that the extreme Kerr throat solution might admit a dual conformal field theory  description
in the spirit of AdS/CFT duality. Almost a decade later, a proposal in \cite{str} initiated extensive investigation
of the Kerr/CFT correspondence which continues today\footnote{By now there is an extensive literature on the subject. For a more complete list of references see e.g. \cite{cms}.}.

When analyzing geometry of a vacuum solution of the Einstein equations, the study of a test particle propagating on the curved background
is instructive in several respects. On the one hand, it facilitates the global analysis of space--time (see e.g. \cite{ge}). On the other hand, a knowledge of first integrals of the geodesic equations helps uncover symmetries of space--time.
For example, the geodesic equations for a massive particle in the Kerr space--time admit
a quadratic first integral which allowed Carter to
integrate the equations by quadratures \cite{car}. This result prompted the authors of \cite{wp} to construct the second rank Killing tensor which is an important geometric characteristic of the Kerr space--time.

As was mentioned in \cite{bh}, a disadvantage of the extreme Kerr throat solution is that it is not supersymmetric.
Note, however, that the corresponding isometry group matches precisely the bosonic subgroup
of $\mathcal{N}=2$ superconformal group in one dimension. At this point it is worth drawing an analogy with the
extreme Reissner--Nordstr\"om black hole.
Near the horizon one reveals the Bertotti--Robinson solution which has the enhanced isometry group $SO(2,1) \times SO(3)$ and is known to be maximally supersymmetric\footnote{Of course, it is to be remembered that the original Reissner--Nordstr\"om solution is supersymmetric.} (for a review see e.g. \cite{moh}). At the same time, $SO(2,1) \times SO(3)$ is the bosonic subgroup of $\mathcal{N}=4$ superconformal group in one dimension. As was shown in \cite{zhou,bgik,gala2}, a massive particle propagating in the Bertotti--Robinson space--time admits $\mathcal{N}=4$ superconformal extension.
Thus, symmetries of the background and the superparticle match.

A relation between bosonic conserved
charges of a test particle propagating on a curved background and Killing vectors characterizing the background geometry is well known.
However, a link between supersymmetry charges and Killing spinors is less well understood.

Motivated by current interest in the extreme Kerr throat geometry,  in this work we construct $\mathcal{N}=2$ supersymmetric
extension of a massive particle moving near the horizon of the extreme Kerr black hole. In the next section
the background geometry is briefly discussed. The second rank Killing tensor is derived by applying the Bardeen and Horowitz limit.
In Section 3 particle dynamics near the extreme Kerr throat is investigated within the framework of the Hamiltonian formalism.
It is shown that the Killing tensor is reducible. Assuming an additional constraint relating the particle and the black hole parameters,
we construct a new class of solutions in elementary functions. Section 4 is devoted to
$\mathcal{N}=2$ supersymmetric extension which is done within the canonical formalism. We summarize our results and discuss possible further developments in Section 5.

\vspace{0.5cm}

\noindent
{\bf 2. Background geometry}\\

Our starting point is the Kerr metric in Boyer--Lindquist coordinates
\bea\label{kerr}
&&
{d s}^2=-\left(\frac{\Delta-a^2 \sin^2{\t}}{\Sigma} \right) dt^2
+\frac{\Sigma}{\D} d r^2+\Sigma d \t^2 +
\left(\frac{{(r^2+a^2)}^2-\Delta a^2 \sin^2{\t}}{\Sigma} \right) \sin^2{\t} d \phi^2
\nonumber\\[2pt]
&&
\qquad \quad
-\frac{2 a  (r^2+a^2-\Delta) \sin^2{\t}}{\Sigma} dt d \phi,
\eea
where
\be
\D=r^2+a^2-2 Mr, \qquad \Sigma=r^2+a^2 \cos^2 {\t}.
\ee
Here $M$ is the mass and $a$ is related to the angular momentum $J$ via $a=J/M$.
In what follows we consider the extreme solution for which $M^2=a^2$. The event horizon is at $r=M$.
Isometries of the Kerr metric include the time translation and rotation around $z$--axis
\be\label{tp}
t'=t+\a, \qquad
\phi'=\phi+\b.
\ee

Apart from the Killing vector fields (\ref{tp}) the Kerr metric (\ref{kerr}) admits the second rank Killing tensor \cite{wp}
which obeys
\be\label{KT}
K_{mn}=K_{nm}, \qquad \nabla_{(n} K_{mp)}=0.
\ee
In Boyer--Lindquist coordinates it reads
\be\label{kt}
K_{mn}=Q_{mn}-r^2 g_{mn},
\ee
where $d s^2=g_{mn} d x^m d x^n$, $x^m=(t,r,\t,\phi)$ and
\be
Q_{mn}=\left(
\begin{array}{cccccc}
-\Delta & 0 & 0 & a \Delta \sin^2{\t} \\
\nonumber\\[2pt]
0 & \frac{\Sigma^2}{\Delta} & 0 & 0\\
\nonumber\\[2pt]
0 & 0 & 0 & 0\\
\nonumber\\[2pt]
a \Delta \sin^2{\t} & 0 & 0 & -a^2 \Delta \sin^4{\t}\\
\end{array}
\right).
\ee

When considering a particle propagating on a curved background with given isometry, each
Killing vector field $\xi^n (x)$ gives rise to the integral of motion $g_{nm} \xi^n \frac{d x^m}{d s}$ which is linear in $\frac{d x^m}{d s}$.
Likewise, the Killing tensor $K_{mn}$ yields the integral of motion $K_{nm} \frac{d x^n}{d s} \frac{d x^m}{d s}$
which is quadratic in $\frac{d x^m}{d s}$. The presence of additional quadratic first integral allowed Carter to
integrate the geodesic equation for a massive particle in Kerr space--time by quadratures \cite{car}.

In order to describe the near horizon geometry, one redefines the coordinates \cite{bh}
\be\label{red}
r \quad \rightarrow \quad M+ \lambda r, \qquad t \quad \rightarrow \quad \frac{t}{\lambda}, \qquad \phi \quad \rightarrow \quad \phi+
\frac{t}{2 M \lambda},
\ee
and then takes the limit $\lambda \rightarrow 0$.
Note that the way the radial coordinate is treated provides a natural definition of the near horizon region.
The last two prescriptions in (\ref{red}) are needed to make $\lim_{\lambda \rightarrow 0 } {d s}^2$ nonsingular
\be\label{kn}
{d s}^2=\left(\frac{1+\cos^2 {\t}}{2} \right) \left[-\frac{r^2}{r_0^2} dt^2+\frac{r_0^2}{r^2} dr^2+r_0^2 d \t^2 \right]+
\frac{2 r_0^2 \sin^2{\t}}{1+\cos^2 {\t} } {\left[d \phi+\frac{r}{r_0^2} dt \right]}^2,
\ee
where $r_0^2=2M^2$. This is a solution of the vacuum Einstein equations \cite{bh}.

Near the throat the isometry group of the metric is enhanced.
In addition to (\ref{tp}) it includes the
dilatation
\be\label{tp1}
t'=t+\g t, \qquad r'=r-\g r,
\ee
and the special conformal transformation
\be\label{sp}
t'=t+\frac 12 (t^2+\frac{r_0^4}{r^2}) \s, \qquad r'=r-tr\s, \qquad \phi'=\phi-\frac{r_0^2}{r} \s.
\ee
Altogether they form $SO(2,1) \times U(1)$ group \cite{bh}.

In the next Section we shall be concerned with the dynamics of a particle propagating near the horizon of the extreme Kerr black hole.
To this end it is instructive to know the explicit form of the corresponding Killing tensor. It
is obtained from (\ref{kt}) by implementing the same limit as above\footnote{The second term in (\ref{kt}) reduces to
the metric and can be discarded.}
\be\label{KT}
K_{nm} d x^n d x^m={\left(\frac{1+\cos^2 {\t}}{2} \right)}^2 \left[-\frac{r^2}{r_0^2} dt^2+\frac{r_0^2}{r^2} dr^2 \right].
\ee
Interestingly enough, up to a conformal factor it coincides
with the $AdS_2$ metric in Poincar\'e coordinates. As will be shown below, this Killing tensor is reducible in the terminology of
\cite{wp} because it can be constructed from the Killing vectors corresponding to $SO(2,1) \times U(1)$ isometry group.

\vspace{0.5cm}

\noindent
{\bf 3. Particle dynamics near extreme Kerr throat}\\

Having fixed the background fields, let us consider the (static gauge) action functional of a relativistic particle
on such a background
\be\label{start}
S=-m\int d t ~\sqrt{\left(\frac{1+\cos^2 {\t}}{2} \right) \left({(r/r_0)}^2-{(r_0/r)}^2 \dot r^2-r_0^2 ~\dot\t^2\right)-\frac{2 r_0^2 \sin^2{\t}}{(1+\cos^2{\t})}
{\left(\dot\phi+r/r_0^2\right)}^2  },
\ee
where $m$ is the mass of the particle.

We choose the Hamiltonian formalism to analyze the particle dynamics. Introducing momenta $(p_r,p_\t,p_\phi)$ canonically
conjugate to the configuration space variables $(r,\t,\phi)$, one builds the Hamiltonian
\be\label{h}
H=\frac{r}{r_0^2}\left(\sqrt{\left(\frac{1+\cos^2 {\t}}{2} \right) {(m r_0)}^2+ {(r p_r)}^2 +p_\t^2+ {\left(\frac{1+\cos^2 {\t}}{2 \sin\t} \right)}^2 p_\phi^2  } -p_\phi\right),
\ee
which corresponds to time translations.
Other isometries of the background metric yield the first integrals
\bea\label{kd}
&&
K=\frac{r_0^2}{r} \left(\sqrt{\left(\frac{1+\cos^2 {\t}}{2} \right) {(m r_0)}^2+ {(r p_r)}^2 +p_\t^2+ {\left(\frac{1+\cos^2 {\t}}{2 \sin\t} \right)}^2 p_\phi^2  } +p_\phi\right)+
\nonumber\\[2pt]
&& \qquad \quad +t^2 H+2tr p_r, \qquad \quad \quad D=tH+r p_r, \qquad \quad \quad \quad  P=p_\phi,
\eea
which generate special conformal transformations, dilatations and rotation around $z$--axis, respectively.
Under the Poisson bracket they form
$so(2,1)\oplus u(1)$ algebra
\be\label{confalg}
\{H,D \}=H, \quad \{H,K \}=2D, \quad \{D,K \} =K.
\ee

The Killing tensor (\ref{KT})
underlies the integral of motion quadratic in the momenta
\be\label{L}
L=\left(\frac{1+\cos^2 {\t}}{2} \right) {(m r_0)}^2+p_\t^2+ {\left(\frac{1+\cos^2 {\t}}{2 \sin\t} \right)}^2 p_\phi^2.
\ee
However, this turns out to be reducible. Computing the Casimir element of $so(2,1)$
one gets
\be
H K-D^2=L-P^2.
\ee
Hence, $L$ is a quadratic combination of the first integrals corresponding to the $SO(2,1)\times U(1)$
symmetry group. Thus, near the horizon the isometry group of the background geometry
is enhanced and the originally independent Killing tensor degenerates.
Note that, in principal, any function of $p_\phi$ and $L$ may represent the $u(1)$ generator in the full symmetry algebra.

Let us integrate the canonical equations of motion
\be
\dot\G=\{\G,H \},
\ee
where $\G=\Big( r,p_r,\t,p_\t,\phi,p_\phi\Big)$ and $\{~,~\}$ stands for the Poisson bracket.
The conserved charges (\ref{h}), (\ref{kd}) and (\ref{L}) allow one to fix the dynamics of the radial pair
\be
r(t)=\frac{r_0^2 E}{\sqrt{a(t)^2+L}-P}, \qquad p_r(t)=\frac{a(t) (\sqrt{a(t)^2+L}-P)}{r_0^2 E }.
\ee
Here $E=H$ is the energy and we abbreviated $a(t)=D-t E$. Note that
$\dot r(t)$ is proportional to $a(t)$ with a positive coefficient. Depending on the initial data, the particle either
goes directly towards the black hole horizon located at $r=0$, or it moves away for some time then
turns at $t=D/E$ and travels back towards $r=0$.

Turning to the canonical pair $(\t,p_\t)$, the first integral (\ref{L}) gives $p_\t$ in terms of $\t$
(for definiteness, we take the positive root)
\be\label{pt}
p_\t=\sqrt{L-\left(\frac{1+\cos^2 {\t}}{2} \right) {(m r_0)}^2-{\left(\frac{1+\cos^2 {\t}}{2 \sin\t} \right)}^2 P^2.}
\ee
Substituting this into the equation of motion for $\t$, one can separate the variables and obtain a solution by quadratures.
It involves elliptic functions.

A special class of solutions is characterized by the additional constraint relating the particle and the black hole parameters
\be
P^2=2 {(mr_0)}^2.
\ee
In this case (\ref{pt}) reduces to
\be
p_\t=
\frac{1}{\sin\t} \sqrt{[L-{(mr_0)}^2]-[L+{(mr_0)}^2]\cos^2\t },
\ee
and the remaining equations of motion for $\t$ and $\phi$ can be integrated in elementary functions
\bea
&&
\cos{\t}=\sqrt{\frac{L-{(mr_0)}^2}{L+{(mr_0)}^2}} \cos{\Big( \sqrt{L+{(mr_0)}^2} (s(t)+\t_0)\Big) },
\nonumber\\[2pt]
&&
\phi(t)=\phi_0-s(t) P+\frac{P}{ \sqrt{2} m r_0} \arctan{\Big[\frac{\sqrt{L+{(m r_0)}^2}}{\sqrt{2} m r_0} \tan{\Big(\sqrt{L+{(m r_0)}^2} (s(t)+\t_0)\Big)} \Big]} -   \nonumber\\[2pt]
&&
\qquad -\frac{P}{16 {\Big(L+{(m r_0)}^2 \Big)}^{3/2}} \Big[ (14 L+10 {(m r_0)}^2) \sqrt{L+{(m r_0)}^2} (s(t)+\t_0)+
\nonumber\\[2pt]
&&
\qquad
+(L-{(m r_0)}^2) \sin{\Big(2 \sqrt{L+{(m r_0)}^2} (s(t)+\t_0) \Big)} \Big]+\ln{\Big(a(t)+\sqrt{L+a(t)^2}\Big)},
\eea
where $\t_0$ and $\phi_0$ are constants of integration and we denoted
\be
s(t)=-\frac{1}{\sqrt{L-P^2}} \Big[\arctan{\Big(\frac{a(t)}{\sqrt{L-P^2}}\Big)} +\arctan{\Big(\frac{a(t) P}{\sqrt{L-P^2} \sqrt{{L+a(t)}^2}}\Big)}\Big].
\ee
Passing further to the Cartesian coordinates one can draw the parametric curve and verify that the orbit looks like a loop which starts
and ends at $r=0$. A similar qualitative behavior was revealed for a massive charged particle moving near the horizon of the extreme
Reissner-Nordstr\"om black hole \cite{gala2}.

\vspace{0.5cm}

\noindent
{\bf 4. $\mathcal{N}=2$ supersymmetric extension}\\

As we have seen above, the global symmetry group of a massive particle propagating near the horizon of the extreme Kerr black hole is
$SO(2,1)\times U(1)$. This is known to be the bosonic subgroup of $\mathcal{N}=2$ superconformal group in
one dimension $SU(1,1|1)$. Our goal in this section is to construct $\mathcal{N}=2$ superconformal extension of the particle model considered above.

Apart from $so(2,1)\oplus u(1)$ generators considered in the previous section,  $\mathcal{N}=2$ superconformal algebra includes the supersymmetry generators
$Q$, $\bar Q$ and the superconformal generators $S$, $\bar S$. Here and in what follows the bar denotes complex conjugation.
In order to accommodate $\mathcal{N}=2$ supersymmetry in the bosonic model governed by the Hamiltonian (\ref{h}), one introduces two fermonic degrees of freedom $\p$ and $\bar\p$
which obey the canonical bracket\footnote{When constructing a supersymmetric generalization, the action functional (\ref{start}) is extended by the fermionic kinetic term
$\int dt \Big(\frac{i}{2} \bar\p \dot \p-\frac{i}{2} \dot{\bar\p} \p \Big)$ as well as other contributions describing boson--fermion couplings. Within the framework of the Hamiltonian formalism there appear fermionic second class constraints,
(\ref{br}) being the Dirac bracket associated to them.}
\be\label{br}
\{\p,\bar\p \}=-i.
\ee

Inspired by the previous study of $\mathcal{N}=4$ superparticle propagating near the horizon of the extreme Reissner-Nordstr\"om black hole \cite{gala2}, we consider the following  supersymmetry generator
\be
Q=\frac{\left( r p_r+i \sqrt{\left(\frac{1+\cos^2 {\t}}{2} \right) {(m r_0)}^2+p_\t^2+ {\left(\frac{1+\cos^2 {\t}}{2 \sin\t} \right)}^2 p_\phi^2-p_\phi^2}  \right)\p }{{\Big[\frac{r_0^2}{2r} \Big(\sqrt{\left(\frac{1+\cos^2 {\t}}{2} \right) {(m r_0)}^2+ {(r p_r)}^2 +p_\t^2+ {\left(\frac{1+\cos^2 {\t}}{2 \sin\t} \right)}^2 p_\phi^2 } +p_\phi \Big)\Big]}^{1/2}}
\ee
and compute the brackets of $Q$ and $\bar Q$
\be
\{Q,\bar Q \}=-2i H, \qquad \{Q,Q \}=0, \qquad \{ \bar Q, \bar Q\}=0.
\ee
These yield the full supersymmetric Hamiltonian
\bea\label{h1}
&&
H=\frac{r}{r_0^2}\left(\sqrt{\left(\frac{1+\cos^2 {\t}}{2} \right) {(m r_0)}^2+ {(r p_r)}^2 +p_\t^2+ {\left(\frac{1+\cos^2 {\t}}{2 \sin\t} \right)}^2 p_\phi^2  } -p_\phi  -
\right.
\nonumber\\[2pt]
&& \qquad \qquad -
\left.
\frac{\sqrt{\left(\frac{1+\cos^2 {\t}}{2} \right) {(m r_0)}^2+p_\t^2+ {\left(\frac{1+\cos^2 {\t}}{2 \sin\t} \right)}^2 p_\phi^2-p_\phi^2} }
{\Big[\sqrt{\left(\frac{1+\cos^2 {\t}}{2} \right) {(m r_0)}^2+ {(r p_r)}^2 +p_\t^2+ {\left(\frac{1+\cos^2 {\t}}{2 \sin\t} \right)}^2 p_\phi^2 } +p_\phi\Big]} \p\bar\p \right).
\eea
Note that its bosonic part correctly reproduces (\ref{h}).  The last term is new and describes the boson--fermion coupling.
That the supersymmetry charges are conserved follows from the Jacobi identities involving $(Q,Q,\bar Q)$ and $(Q,\bar Q,\bar Q)$.

The generators of dilatations and special conformal transformations are constructed as
in (\ref{kd}), with $H$ being now the supersymmetric Hamiltonian (\ref{h1}). Altogether they prove to
form the conformal algebra (\ref{confalg}).

In order to find the generator of superconformal transformations $S$, it is sufficient to compute the bracket of $K$ and $Q$
\be
\{K,Q \}=S,
\ee
which yields
\be
S=-t Q-{\Big[\frac{2 r_0^2}{r} \left(\sqrt{\left(\frac{1+\cos^2 {\t}}{2} \right) {(m r_0)}^2+ {(r p_r)}^2 +p_\t^2+ {\left(\frac{1+\cos^2 {\t}}{2 \sin\t} \right)}^2 p_\phi^2 } +p_\phi \right)\Big]}^{1/2} \p.
\ee
Calculating further the bracket of $Q$ and $\bar S$
\be
\{Q,\bar S \}=2i (D+i J),
\ee
one finds the $u(1)$ $R$--symmetry generator $J$ in the full $\mathcal{N}=2$ superconformal algebra
\be
J=\sqrt{\left(\frac{1+\cos^2 {\t}}{2} \right) {(m r_0)}^2+p_\t^2+ {\left(\frac{1+\cos^2 {\t}}{2 \sin\t} \right)}^2 p_\phi^2-p_\phi^2}+\frac 12 \p\bar \p.
\ee
The rest of the algebra reads
\begin{align}
&
\{D,Q \}=-\frac 12 Q, && \{H,S \}=-Q, && \{D,S \}=\frac 12 S,
\nonumber\\[2pt]
&
\{S,\bar S \}=-2iK, && \{J,Q \}=-\frac i2 Q, && \{J,S \}=-\frac i2 S,
\end{align}
plus complex conjugate relations. The vanishing brackets were omitted.
Finally, it is readily verified that $S$, $\bar S$ and $J$ are conserved in time.

Note that $P=p_\phi$ is also a first integral which commutes with all other generators from $\mathcal{N}=2$ superalgebra.
Thus, the full symmetry group of the superparticle is
$SU(1,1|1) \times U(1)$. When taking the bosonic limit, the generator $J$ becomes the
square root of the Casimir element of $so(2,1)$ which can be discarded\footnote{We thank G. Comp\'ere for pointing this out to us.}.
This is to be contrasted with the $\mathcal{N}=4$ superparticle in the Bertotti--Robinson space--time \cite{bgik,gala2} where
no extra $U(1)$ factor appear.
As to the quadratic first integral $L$ defined in (\ref{L}), like in the bosonic case, it turns out to be  reducible
\be
\frac{1}{2} (H K-D^2+J^2)=L-P^2.
\ee

To summarize, we have constructed a superparticle model invariant under $SU(1,1|1) \times U(1)$ supergroup, which in the bosonic limit
reduces to the particle moving near the extreme Kerr throat.

\vspace{0.5cm}

\noindent
{\bf 5. Conclusion}

\vspace{0.5cm}

In the present paper we
studied the dynamics of a massive particle moving near the horizon of the extreme Kerr black hole within the framework of the Hamiltonian formalism.
It was shown that for the extreme Kerr throat geometry the second rank Killing tensor is reducible and up to a conformal factor
it coincides with the $AdS_2$ metric in Poincar\'e coordinates.
For a special choice of particle parameters a new class of solutions was built which was given in terms of elementary functions.
$\mathcal{N}=2$ supersymmetric extension of the particle model was constructed within the canonical formalism.

Turning to possible further developments, the most urgent question is to clarify geometric origin of the $\mathcal{N}=2$
supersymmetry charges constructed above. To this end it is important to build a super 0--brane action in curved space--time for which the formulation
in Section 4 is a gauge fixed Hamiltonian version. Then it is interesting to generalize the present analysis to the
case of the near horizon geometry of the extreme Kerr-Newman black hole \cite{bh,hor}.
Higher dimensional generalizations are also worth studying.

As is well known, for a particle model on a curved background admitting Killing--Yano tensors a new type of supersymmetry charges can be constructed \cite{grh} (for some recent developments see \cite{hh}). In particular, their Poisson brackets yield Killing tensors. It is interesting to construct such new supercharges for the $\mathcal{N}=2$ superparticle presented in this work and compare them with the conventional supercharges.

$\mathcal{N}=4$ superparticle in the Bertotti--Robinson space--time is related to conventional conformal mechanics by a canonical transformation of variables \cite{bgik,gala2}. Such a transformation supports a black hole interpretation of the conformal mechanics given in
\cite{kt}. A conformal mechanics counterpart of the $\mathcal{N}=2$ superparticle considered in this work is also worthy to investigate.

Finally, it is interesting to study whether the particle moving near the extreme Kerr throat can be obtained by the Hamiltonian reduction from
a larger higher dimensional system in the spirit of \cite{an}.

\vspace{0.5cm}

\noindent{\bf Acknowledgements}\\

\noindent
We thank  Geoffrey Comp\'ere, Gary Gibbons, Peter Horv\'athy, Armen Nersessian for useful comments and
Evgeny Ivanov for interesting discussions.
This work was supported by the Dynasty Foundation and in part by grants RFBR 09-02-91349, LSS 3558.2010.2, "Kadry" P691.

\end{document}